# On the phase behaviour of pure PSPC and PEGylated multi-component lipid and their interaction with Paclitaxel: An all-atom MD study

Prantar Dutta[1,#], Debabrata Pramanik[1,#], Jayant K. Singh[1,2]*

[1]Department of Chemical Engineering, Indian Institute of Technology Kanpur, Kanpur 208016, India.

[2]Prescience Insilico Private Limited,

Old Madras Road, Bangalore 560049, India

[#]Equal Contribution,

*Correspondence to: Jayant K. Singh

Phone: 91-512-259 6141 (Office), Fax: 91-512-259 0104, Email: jayantks@iitk.ac.in

## Abstract

The study of the structural behaviour of pure and multi-component lipids at various temperatures and the interaction of these multi-component lipids with pharmaceutically important drugs carry huge importance. Here, we investigated the phase behaviour of the pure PSPC (1-palmitoyl-2-stearoyl-sn-glycero-3-phosphocholine), and multicomponent PSPC and DSPE-PEG$_{2000}$(1,2-distearoyl-sn-glycero-3-phosphoethanolamine-N[amino(polyethylene glycol)-2000]) membranes at seven different temperatures ranging from 280 K to 360 K, and calculated their structural properties. We observe a transition from the gel phase to the liquid crystalline phase between 320 K and 330 K in agreement with experimental reports for pure PSPC. PSPC remained in the tilted gel phase $L_{\beta'}$ at 320 K and 310 K, entered the "mixed ordered" domain with a partially interdigitated region at 300 K, and finally formed the sub gel phase at 280 K. We studied the self-assembly for the multicomponent PSPC and DSPE-PEG$_{2000}$ membranes and found the coexistence of ordered and disordered phases at 320 K. In comparison to the pure PSPC, for multicomponent system, this transition was gradual, and a complete liquid crystalline to gel phase transformation occurred between 320 K and 310 K. We further studied the interaction of Paclitaxel with pure PSPC and PEGylated multicomponent lipid bilayers using umbrella sampling technique and observed PEG promotes the interaction of Paclitaxel with the later one in comparison to the former. Above the bilayer transition temperature, Paclitaxel interacts more with the bilayer and enters inside the bilayer easily for both systems. Understanding of structural and interaction behaviour of the PEGylated multicomponent lipid bilayers with Paclitaxel will help explore Paclitaxel based drug applications in the future.

# 1. Introduction

Liposomes are artificial spherical vesicles composed of phospholipids and cholesterol, with diameters ranging from tens of nanometres to a few microns. They generally consist of a lipid bilayer surrounding an aqueous core and hence can be loaded with both hydrophobic and hydrophilic solutes. Due to their minuscule size, amphiphilic nature, high biocompatibility, stabilizing effect on drugs, and non-toxicity, liposomes have been widely investigated for drug delivery applications.[1,2] These nanocarriers deliver the therapeutic molecules at the target site by fusing its lipid bilayer with the bilayers of the cell membrane, followed by diffusion of the liposomal content into the cell interiors. But this process is non-spontaneous and highly complex. Targeted delivery can be achieved by designing stimuli-responsive systems to release the encapsulated drug at specific physicochemical conditions. For example, drug release in thermosensitive liposomes is triggered by temperature. Lipid bilayers are known to undergo a transition from ordered to disordered phase at a specific temperature. The molecular arrangement undergoes an abrupt change at the phase transition, and the liposomal contents are released. The liposome needs to be designed in such a way that the phase transition temperature is slightly above physiological temperature. Then, applying localized hyperthermia can trigger drug release at the target site. However, if the transition temperature is considerably higher than normal body temperature, the surrounding tissues may get damaged. Lipids are often functionalized with polymer chains like Polyethylene Glycol (PEG), and the corresponding systems are said to be "PEGylated". PEGylation increases the blood circulation time after intravenous injection of liposomes by providing stealth from the immune response.[3,4] Protein absorption on the liposome surface is hindered sterically, which in turn inhibits uptake by macrophages.

Liposome based formulations are extensively used in cancer treatment because conventional chemotherapy suffers from several limitations like non-specific targeting, high-dose requirements, undesirable side effects, low bioavailability, and development of multiple drug resistance. Paclitaxel (PTX), also known by its brand name Taxol®, is a potent anti-tumour molecule that is widely used against a variety of cancers, including ovarian, breast, lung, and pancreatic. Although, traditionally, PTX was extracted from yew trees only, environmentally-benign routes of production from microorganisms have been developed in recent years.[5] Albumin-bound PTX nanoparticle, called Abraxane® is an injectable chemotherapeutic medication for treating metastatic conditions.[6] Nonetheless, its agonizing side effects like neuropathy, a decrease in white blood cells, shortness of breath, allergic reactions, and acute infections drive the pursuit of alternatives. In search of alternatives, contemporary experimental studies of liposomal PTX presents promising evidence of superior drug delivery systems.[7–10]

Optimum design of thermosensitive liposomes is based on understanding the phase behaviour of the constituent lipids. Phospholipid bilayers exhibit different phases depending on temperature, pressure, and composition, each with its characteristic structural features. The primary phase transition is from the gel phase at low temperatures to the liquid crystalline phase at high temperatures. The lamellar gel phase can be of two types— $L_\beta$ and $L_{\beta'}$, both with ordered lipid tails. The tails in $L_\beta$, typical of phosphatidylethanolamine (PE) lipids, are almost parallel to the bilayer normal while the tails in $L_{\beta'}$, observed in phosphatidylcholine (PC) lipids, are slightly tilted. Cooling below the gel phase leads to the sub gel ($L_c$) phase in some lipids, where the hydrocarbon tails are highly ordered and tilted. At higher temperatures, liquid crystalline phase ($L_\alpha$) is observed, where the tails are disordered and do not display any tilt[15]. The transition from $L_{\beta'}$ to $L_\alpha$ often proceeds through an intermediate ripple phase ($P_{\beta'}$) for PC lipids where the bilayer is periodically corrugated and both ordered and disordered nanodomains coexist. The interdigitated gel phase, $L_{\beta I}$, is also observed in a few PC lipids. Here, some of the lipid tails of the two leaflets substantially overlap, primarily due to gaps between the headgroups.

Experimental techniques like DSC, XRD, NMR spectroscopy, FTIR and fluorescence measurements have been employed for a long time to study the phase behaviour of lipid bilayers.[11–14] In the last couple of decades, molecular simulations emerged as a powerful tool to investigate nanoscopic details of lipid membranes. Leekumjorn and Sum computed a wide range of structural properties during the phase transition of DPPC and DPPE bilayers using atomistic molecular dynamics (MD).[15] In their simulations, three distinct structures of DPPC were identified: (i) tilted lipids and overlap between the tails of the two leaflets below $T_m$, (ii) disordered fluid phase above $T_m$, and (iii) a transitional phase near $T_m$ with elevated bilayer thickness and area per lipid. DPPE neither showed lipid tail overlap, nor an intermediate phase. Lecithin lipid bilayers, studied by de Vries et al., rippled upon cooling from the $L_\alpha$ phase.[16] They observed a gel phase and a gel-like phase with interdigitated lipid tails connected by a kink of disordered lipids. Khakbaz and Klauda predicted the $T_m$ of DPPC and DMPC using all-atom MD with high accuracy.[17] Furthermore, their work quantified the structure of the $P_{\beta'}$ phase of DMPC and validated the CHARMM36 force-field for phase transition. Coarse-grained (CG) models are also applied to simulate lipid bilayers[18–20], but they are often unable to capture minute structural details important in phase transition. A compilation of simulation studies of lipid domains can be found in the review by Bennett and Tieleman.[21]

The structure of PEG-conjugated lipid bilayers is of significant interest to researchers because of their importance in drug delivery. Early experiments with DSPC and DSPE-PEG liposomes by Kenworthy et al. provided profound insights into the structure and phase behaviour of PEGylated membranes.[22] PEG chains of molecular weights in the range of 350 to 5000 were considered for preparing systems with 0-60 mol % DSPE-PEG. $T_m$ values varied from 55 °C to 64 °C depending on the PEG size and concentration. XRD showed the formation of $L_{\beta'}$ at

PEG concentrations of 10 % and below for all molecular weights, untilted $L_{\beta I}$ for high concentrations but low molecular weights, and $L_\beta$ for high concentrations and high molecular weights. Johnsson and Edwards conducted cryo-TEM experiments of DOPE with DSPE-PEG to show that the PEG size affects the bilayer phase behaviour.[23] Atomistic MD and Langmuir monolayer film experiments of PEGylated liposomes by Stepniewski et al. in physiological conditions demonstrated that the PEG chains penetrate the bilayer core in the fluid phase but remain exposed to the surrounding solution in the gel phase.[24] Lee and Pastor self-assembled pure and PEG-grafted DPPC using the CG Martini force-field to form liposomes, micelles, and bicelles.[25] With increasing PEGylated lipid concentration, the aggregate size decreased for all the assemblies. Shinoda et al. reported similar findings for pure and PEGylated DMPC system.[26] For liposome design, it is imperative to understand the interactions of drug molecules with PEGylated membranes. Lu et al. simulated the interactions of PTX with model membranes under the effect of shock waves and nanobubbles.[27] Kang and Loverde performed MD simulations and free energy calculations of PTX transfer from water to a POPC membrane wherein the concentration of drug in the membrane affected the partitioning and free energy profile.[28] However, there are no studies in the literature which describe the phase behaviour of the PEGylated bilayers and their interaction with the PTX. Here, using all-atom MD we study the phase behaviour of pure PSPC (1-palmitoyl-2-stearoyl-sn-glycero-3-phosphocholine), and multicomponent PSPC and DSPE-PEG$_{2000}$ (1,2-distearoyl-sn-glycero-3-phosphoethanolamine-N[amino(polyethylene glycol)-2000]) membranes at temperatures ranging from 280 K to 360 K, and the interaction of the PTX with these bilayers using umbrella sampling simulations.

## Model and Simulation Details

The PSPC and DSPE-PEG$_{2000}$ molecules were modeled in Materials Studio.[29] The molecular geometries were determined by B3LYP Density Functional method using Gaussian09 with 6-

31G(d,p) basis sets.[30] The partial atomic charges were calculated with the CHELPG scheme implemented in Gaussian09. We took these optimized structures for MD simulations. The CHARMM36 all-atom force field described the molecular interactions.[31–34] The structure and forcefield of PTX were obtained from a previous work by Kulkarni and coworkers.[35] TIP3P water was used to solvate all the simulation boxes.[36] As DSPE is an anionic lipid, the excess charge in the system was neutralized by $Na^+$ counter ions. To improve computational efficiency, non-bonded interactions were cut off beyond a radius of 1.2 nm. The LJ forces were smoothly decreased to zero between 1 nm and 1.2 nm. The Particle Mesh Ewald (PME) method accounted for the long-range contribution of the electrostatics.[37] The bond lengths were constrained using the LINCS algorithm.[38] The temperature and pressure were controlled by the Parrinello-Bussi velocity rescale thermostat[39] with 0.1 ps coupling time constant and Parrinello-Rahman barostat[40] with 2 ps coupling time constant, respectively. The steepest descent algorithm minimized the potential energy of starting structures prior to the simulations. The equations of motion were solved by the leapfrog algorithm with a time step of 2 fs. Periodic boundary conditions were implemented in x, y, and z directions. All MD simulations were performed using the GROMACS 2018.4 code.[41]

For the pure PSPC simulations, the initial configurations were generated in Packmol.[42] 128 lipids, divided equally between the two leaflets, were packed in a 7 nm × 7 nm × 5.6 nm unit cell. The arrangement resembled a bilayer structure, with the normal in the z-direction. Water slabs of the same planar dimensions as that of the cell and 3 nm thickness were added at the top and bottom of the bilayer to solvate the structure. The entire system was put in a 7 nm × 7 nm × 13 nm box, and energy minimized. We simulated the system at seven temperatures: 280 K, 300 K, 310 K, 320 K, 330 K, 340 K, and 360 K. 10 ns NVT equilibration, 10 ns NPT equilibration, and 200 ns production were performed for all the cases. The barostat was coupled semi-isotropically during the production runs with the pressure set at 1 bar. All analyses were performed with the last 100 ns of simulation trajectory. In Table 1, we mentioned details of the equilibrium simulations performed in the study.

For the multicomponent system, we self-assembled a stable bilayer from a random configuration. 128 PSPC and 8 DSPE-PEG2000 molecules were arbitrarily placed in a 15 nm × 15 nm × 15 nm box and solvated. We minimized the potential energy and then equilibrated in NVT and NPT ensembles for 10 ns and 20 ns, respectively. The system was simulated for 200 ns at 360 K and 1 bar with semi-isotropic pressure coupling, leading to a bilayer formation. The water molecules were removed from the structure, and the bilayer was re-solvated in such a way as to prevent the PEG chains of the two leaflets from interacting with each other because of periodic boundaries. The same sequence of steps— energy minimization, 10 ns NVT equilibration, 10 ns NPT equilibration, and 200 ns production run— was repeated at 360 K. Simulated annealing of the stable bilayer was performed to generate structures at 280 K, 300 K, 310 K, 320 K, 330 K, and 340 K. A linear cooling rate of 5 K/ns was considered for the annealing process. Finally, NVT and NPT equilibrations of 10 ns each, followed by 200 ns production run was performed for all the temperature points. Similar to the pure PSPC systems, the last 100 ns of the trajectories were analysed for property calculations.

Thus from brute force MD simulations we elucidated various structural quantities for the pure PSPC and PEGylated PSPC systems at different temperatures. Further to understand the interaction behaviour between the PTX drug with rest of the lipid systems from a quantitative perspective, we performed umbrella sampling (US) based enhanced sampling simulations.[43–47] In the US simulations we choose centre of mass (COM) distance between the heavy atoms of the ligand and bilayer plane as the reaction coordinate to bias the system. The distance was chosen in the plane perpendicular to the bilayer plane. Initially the ligand is placed in bulk in the aqueous medium and gradually we decrease the distance to bring the ligand inside the bilayer though the interface.

We performed umbrella sampling simulations for four systems at two temperatures (310 K and 330 K). For each system, we simulated first steered MD simulations by pulling the drug molecule from bulk to the interior region. We generated configurations having each window 0.1 nm apart for the US simulations for each system. These configurations have been taken as the initial configurations and then applied harmonic restrained potential for the US simulations. A harmonic potential with a force constant (k) of 1000 kJ.mol$^{-1}$.nm$^{-2}$ was used to bias the system in the US simulations. The initial and final distances for the reaction coordinate were taken as 5 and 0 nm respectively, and in between 50 umbrella sampling windows were simulated. Each US windows were simulated for 10 ns and thus for each system we simulated a total of 500 ns or more simulation time. Using GROMACS suite we performed US simulations and later used WHAM (weighted histogram analysis method)[48,49] to extract the potential of mean force (PMF).[50] We performed three independent US simulations for each system to have statistically reliable results. Similar simulations protocol have been employed for all the four systems and the PMFs have been extracted. In Table 2 we mentioned details about the systems for the US simulations.

## 3. Results and Discussions

### 3.1 Pure PSPC Bilayer

PSPC has an experimental main phase transition value of around 49 °C at atmospheric pressure[46], which acts as a reference to validate our findings. We first considered two fundamental structural properties of lipid bilayers— area per lipid (APL) and bilayer thickness, at different temperatures. APL was calculated by dividing the box area in the x-y plane by the number of lipids in each leaflet (64 in this case). We plotted the fluctuations in APL during the last 100 ns of simulation for each temperature and observed minimal drift in the mean value (Figure S1 in Supplementary Information). Hence, the bilayers were assumed to be sufficiently equilibrated and suitable for property computations. The fluctuations were quite pronounced at 330 K and above, as compared to lower temperatures. Figure 1 (a) shows the average APL as a function of temperature. We took block averages by dividing the APL data into five segments of 20 ns simulation time each, and reported the average APL as the mean of those block averages. The error bars denote the maximum positive and negative deviations of the block averages from the mean. In agreement with the experiment, the APL-temperature curve shows a phase transition between 320 K and 330 K. The bilayer enters the liquid crystalline phase above $T_m$, and the APL continues to increase linearly with an increase in temperature. However, the picture is more complicated below $T_m$ as no clear correlation between APL and temperature is evident. For determining the bilayer thickness, the density profile of the headgroup phosphorous atoms of PSPC was plotted as a function of the distance from the bilayer midplane (Figure S2 in Supplementary Information). Sharp peaks were observed at 330 K, 340 K, and 360 K. The peaks get broader as the temperature is lowered. Two distinct peaks on each side of the bilayer midplane was seen at 300 K. Bilayer thickness was calculated as the distance between the density profile peaks. In the case of 300 K, the peaks farther away from the midplane were considered for the calculations. The variation of bilayer thickness with temperature is shown in Figure 1 (b). The mean values and error bars were calculated by taking block averages, as in the case of APL. Similar to the APL curve, we see a transition between 320 K and 330 K. The thickness decreases almost linearly with temperature rise for the liquid crystalline phase. No interdependence trend between bilayer thickness and temperature is apparent below phase transition.

To understand and rationalize these behaviours, we looked at the bilayer molecular arrangement at different temperatures. Figure 2 presents the snapshots of the PSPC bilayer above $T_m$. The disordered tails permit the lipids to diffuse around in their respective leaflets. As the headgroups aren't stacked against each other, the available cross-section area (and consequently APL) is high. The curving of the hydrocarbon tails at elevated temperatures leads to compaction along the normal direction, thus reducing the bilayer thickness. The APL and bilayer thickness variation trend with the temperature of PSPC in the $L_\alpha$ phase is similar to the findings for other PC bilayers. There is no visually observable qualitative difference between the three structures shown in the figure. Figure 3 shows the bilayer structure below the transition temperature, as seen from x-z and y-z planes. The $L_{\beta'}$ gel phase, with tilted tails and no interdigitation, is observed at 320 K and 310 K. The tilt angle increases going from 320 K to 310 K, which in turn reduces the bilayer thickness. At 300 K, two different domains can be visualized: (i) a highly ordered gel phase with stretched and minimally tilted tails but with no interdigitation and (ii) an interdigitated gel phase with similar highly ordered tails. The adjacent locations of these domains indicate rippling in the bilayer structure, as is evident from Figure 3 (d). However, our system size is not large enough to conclusively capture any rippled phase. We refer to this structure as the "mixed ordered" domain. The interdigitation is presumably due to the asymmetric chain lengths of the two hydrocarbon tails. The bilayer thickness is high at 300 K due to the swelled region with no overlap. Compared to other gel phase structures, the APL increases as well because the close packing of the headgroups gets disrupted due to rippling. At 280 K, the surface of the bilayer shows unusual defects with slight overlap regions between the leaflets. Some of the headgroups appeared to have melted. The lipid tails are slightly tilted, typical of the sub gel phase in PC bilayers. We simulated multiple replicas at 280 K, 300 K, and 310 K to confirm the bilayer structures and obtained similar results. Hence, the APL and bilayer thickness findings can be explicated by the underlying molecular details.

The deuterium order parameter ($S_{CD}$) was calculated along the acyl chains for PSPC with respect to the bilayer normal, as shown in Figure 4. It is a measure of lipid orientation in the bilayer and can capture phase transitions. In accordance with previous results, a significant difference is noted between the curves from 330 K to 320 K for both the acyl chains. In the $L_\alpha$ phase, the order decreases slightly as the temperature goes up because of the enhanced fluidity. Going from 320 K to 310 K, $S_{CD}$ falls due to a greater tilt angle of the lipid tails with respect to the bilayer normal at 310 K. The "mixed ordered" domain at 300 K is highly ordered as the tails are straightened, making them aligned with the bilayer normal. A slight tilt appears at 280 K (Figure 4 (a) and (b)), causing a reduction of $S_{CD}$. The dependence of order on temperature is similar for the two acyl chains. Figure S3 presents the variation of the average order parameter ($S_{D,avg}$) with temperature. $S_{D,avg}$ is considered as the mean of $S_{CD}$ values for all atoms in the two hydrocarbon tails.

We also calculated the torsional distribution of the lipid hydrocarbon tails, as shown in Figure 5. Dihedral angles of around ± 180° correspond to a *trans* conformer while the angles of *gauche* conformers range from ± 60° to ± 120°. The proportion of *trans* is highest at 280 K and decreases with temperature rise. The opposite trend is observed for *gauche*. When the lipids are highly ordered, the *trans* conformer is favoured energetically. The increased flexibility in the $L_\alpha$ phase produces more *gauche* conformers. We see a gap corresponding to the phase transition between 320 K and 330 K for both regions of the distribution curve.

## 3.2 PSPC and DSPE-PEG Mixed Bilayer

We self-assembled the multicomponent bilayer at 360 K and 1 bar from a random configuration following the series of steps described in section 2. Figure 6 shows the initial and final structure of the self-assembly simulation. The PSPC molecules are found to be unevenly distributed in the bilayer, with 58 lipids in the top leaflet at 70 in the bottom. In contrast, both the leaflets have four DSPE-PEG$_{2000}$ each. The PEG chains are coiled and exposed to water. No particular preference for any specific molecule or group was observed for the Na$^+$ ions, all of which remained in the aqueous medium. The disordered hydrocarbon tails of the lipids represent the liquid crystalline phase. A series of structures were generated at temperatures ranging from 280 K to 340 K by cooling the bilayer from 360 K, as shown in Figure 7.

The bilayer continues to be in the fluid $L_\alpha$ phase at 340 K and 330 K. This is evident from the structure of the lipid tails in the snapshots. PSPC behaves in the same way in this PEGylated multicomponent system as it does for the pure bilayer at temperatures above phase transition. At 320 K, we find two distinct domains: (i) an ordered region with slightly tilted and completely interdigitated lipid tails and (ii) a disordered region almost similar to the liquid crystalline phase. The disordered domains occur in the vicinity of the PEGylated DSPE, especially where the PEG chains penetrate the bilayer surface. The resultant structure is an intermediate between gel and liquid crystalline phase, and henceforth referred to as the "mixed" domain. Further cooling of the bilayer leads to a highly ordered gel phase with stretched and interdigitated lipid tails. Thus, for the multicomponent system, the transition occurs gradually. Although $T_m$ is above 320 K for the PSPC bilayer, complete transition in the PEGylated bilayer takes place between 310 K and 320 K. Comparing with the structures shown in Figure 3, the effect of PEG on the gel phase of the membrane is seen to be two-fold. Firstly, it reduces the tilt observed in pure PSPC membrane (characteristic of $L_{\beta'}$ phase) at 310 K and 320 K. Secondly, it induces the formation of the interdigitated gel phase $L_{\beta I}$. The cause of interdigitation can be traced back to the self-assembly process. The membrane aggregated in such a way that the top leaflet has fewer number of lipids than the bottom, but both occupied the same lateral area. As a result, the area available per headgroup is quite higher in the top leaflet. Consequently, there is void formation within the lipid packing. When the bilayer is cooled below $T_m$, the hydrocarbon tails of the lipids are stretched and simply enter the voids to occupy the energetically favourable state. We also observe localized disruptions in the ordered molecular arrangement near the bilayer headgroups at 300 K and 310 K due to PEG penetration. Cooling the membrane to 280

K forms the sub gel phase with tilted tails and substantial overlap between the two leaflets. Overall, the mixed bilayer of PSPC and DSPE-PEG$_{2000}$ has a radically different phase behaviour than pure PSPC.

We computed the APL for the top and bottom leaflet separately, as shown in Figure 8 (a) and (b). As expected, it is considerably higher for the top leaflet than the bottom one. The APL increases with temperature, but unlike the PSPC bilayer, no drastic change is observed at transition. For both leaflets, two separate regions can be distinguished— (i) 280 K < T < 310 K and (ii) 320 K < T < 360 K. Both regions are linear, but the slope changes between 310 K and 320 K. Because the $T_m$ of DSPE is around 74 °C, the multicomponent system is expected to have a higher $T_m$ than pure PSPC. But the opposite happens due to PEG conjugation. The APL results agree with the molecular structures discussed earlier. Even for the lower leaflet, the APL is slightly higher than the single-component bilayer in the liquid crystalline phase and notably higher below $T_m$. The heterogeneity of the membrane and the partial ingress of the coiled PEG chains into its hydrophobic core produce this behaviour. The headgroups are no longer densely packed, facilitating the lipid tail overlap between the two leaflets. Figure 8 (c) presents the dependence of bilayer thickness of the multicomponent membrane on temperature. Contrary to pure PSPC bilayer, the thickness here is higher for the liquid crystalline phase than the gel phase due to interdigitation. The two layers fuse into each other below $T_m$, which brings down the distance between the headgroup phosphorous atoms. A sharp rise is noted between 310 K and 330 K, indicating two distinct phases. The "mixed" domain is responsible for the behaviour at 320 K, which shows an intermediate bilayer thickness. At 280 K, the thickness further decreases due to tilt in the lipid tails. In the $L_\alpha$ phase, the bilayer thickness is slightly lower for the multicomponent system compared to the PSPC membrane. In the gel phase, it is drastically different for the two cases.

We also calculated the deuterium order parameter ($S_{CD}$) for the two hydrocarbon tails of PSPC in the multicomponent bilayer, as shown in Figure 9. The curves of the three systems in the $L_\alpha$ phase (330 K, 340 K, and 360 K) almost overlap each other for both the acyl chains. Similar to the observations of bilayer thickness, there is a pronounced leap going from 330 K to 310 K, while 320 K occupies a transitional position. At 280 K, the tilt of the lipid tails with respect to the bilayer normal decreases the $S_{CD}$. Figure 10 shows the dihedral distribution of the C-C-C-C dihedral of the lipid tails. Matching with the previous findings for pure PSPC, the proportion of *gauche* conformers increases, and that of *trans* conformers decreases as temperature rises. A phase transition between 320 K and 330 K can be detected from the plots. Combining the observations of all the four structural quantities calculated for the PEGylated multicomponent bilayer, we infer that the properties of the "mixed" bilayer at 320 K resemble the liquid crystalline phase more than the gel phase. Although there is no sharp transition like pure PSPC, complete transformation to the gel phase occurs between 310 K and 320 K. Hence, PEG brings down the $T_m$ of the membrane and distorts the typical gel phase structure.

## 3.3 Interaction of the Bilayers with Paclitaxel

In Figure S4 to S7 we showed the histogram plots of the umbrella sampling windows for pure PSPC and PEGylated systems at two temperatures (310 K and 330 K). Here we have shown histograms only for one independent simulations out of three for respective systems. A good overlap between the successive windows along the reaction coordinate for the US simulations show reliability of the PMF calculations. We calculated average PMFs averaging over independent probabilities and the averaged PMFs are shown in Figure 11. Figure 11 (a) presents the PMFs for pure PSPC and PEGylated systems at 310 K, and Figure 11 (b) shows the PMFs for pure PSPC and PEGylated systems at 330 K. To have the microscopic understanding at the all-atom level, instantaneous snapshots have been added in the PMF plots at the bulk (non-interacting region), at the interface of the water-bilayer, and at the minimum of the PMFs. From the instantaneous snapshots (Figure 11 (a)), we see that for the case of pure PSPC, it is at the tilted gel phase at 310 K and shows high repulsion at the interface of the bilayer. As a result, the PTX cannot enter the bilayer easily. However, with the introduction of the PEG in the system, the gel phase goes to the inter-digitated gel phase. This opens up space for the PTX to enter inside the bilayer, as shown in the Figure S8 (a). The PMF plots (Figure 11 (a) inset) show that with the introduction of the PEG in the system, the barrier decreases in comparison to the pure PSPC. The quantitative values of the dissociation barriers have been shown in Table 3. Figure 11(b) shows the PMFs, at 330 K, along with instantaneous snapshots of the microscopic conformations at different positions of the potential well. From the PMF plots, we see that at 330 K (at the liquid crystalline phase) for pure PSPC, PTX interacts more with the bilayer interface (as shown by the instantaneous snapshots) due to increased entropy in comparison to that in 310 K. We see a dissociation barrier of 0.30 (0.323) kcal/mol and interface barrier of 0.52 (0.145) kcal/mol for the pure PSPC. With the introduction of the PEG in the system, the bilayer interface opens up for PTX as shown in Figure S8 (b), and as a result PTX interacts more with the bilayer than pure PSPC. From the PMF plots we see a dissociation barrier of 0.59 (0.141) kcal/mol and interface barrier of 0.73 (0.032) kcal/mol for PEGylated bilayer system (see Table 3). Thus with the introduction of the PEG in the system, the DSPE-PEG molecules open up more spaces at the bilayer interface compared to the pure PSPC, which promotes the interactions of the PTX with the lipid bilayer and enters inside the bilayer. The opening of the space at the interface of the bilayer increases the average area per lipids which is shown by the quantitative estimation of the structural properties calculations with increasing temperatures as seen in Figure 1 (a) for pure PSPC, and in Figure 8 (a) and (b) for PEGylated multicomponent bilayers. Thus the calculated structural properties show a good correlation with that of the interaction pictures for the pure lipid bilayer and PEGylated multicomponent bilayers.

## 4. Conclusions

In this work, using all-atom MD simulations, we studied the phase behaviour of the pure PSPC and multicomponent PSPC and DSPE-PEG$_{2000}$ membranes. PSPC was simulated at seven different temperatures ranging from 280 K to 360 K. In agreement with experimental results, the transition from the gel phase to the liquid crystalline phase occurred between 320 K and 330 K. PSPC remained in the tilted gel phase $L_{\beta'}$ at 320 K and 310 K, entered the "mixed ordered" domain with a partially interdigitated region at 300 K, and finally formed the sub gel phase at 280 K. The self-assembly of the multicomponent system at 360 K generated an asymmetric bilayer with fewer lipids in the top leaflet. The membrane was annealed to obtains configurations at the same temperature points as that of pure PSPC. Uneven lipid distribution between the leaflets induced voids in the bilayer, which led to interdigitation in the gel phase. We found the coexistence of ordered and disordered phases at 320 K and characterized the structure as a "mixed" domain. The phase transition was gradual, and a complete liquid crystalline to gel phase transformation occurred between 320 K and 310 K. We calculated averaged PMFs, from the umbrella sampling simulations, for pure PSPC and PEGylated bilayers at 310 K and 330 K. From the PMF calculations, we show that at 310 K with the introduction of the PEG in the system, it goes from the tilted gel phase to the interdigitated gel phase and the PTX interacts more for PEGylated bilayers. At 330 K at the liquid crystalline phase with the introduction of the PEG in the system, instantaneous snapshots and a closer look at the interface conformations show that with the interaction of the PEG, there is opening of spaces at the interfaces for the PTX and thus PTX interacts more with the multicomponent lipid bilayers than in pure PSPC even at 330 K. Thus, these studies explore the structural behaviour of the pure PSPC and PEGylated multicomponent lipid bilayers at different temperatures and microscopic behaviour and the quantitative values of the interactions of the anti-tumour drug

Paclitaxel with that of lipid bilayers at the gel phase and liquid crystalline phase of the bilayer. These studies are envisaged to have a large impact on the understanding and developing Paclitaxel based drug application in the future. The effect of the PEG concentration on the phase behaviour of the multicomponent lipid bilayers and on the interaction of the PTX with the bilayers will be a context for future studies. Also, the pore opening mechanism at the bilayer interface by the addition of the PEG in the system is of great importance to explore in a later study.

## Conflicts of interest

There are no conflicts of interest to declare.

## Acknowledgements

DP thanks IIT Kanpur for generous financial assistance.

## Supporting Information

Area per lipid profile for pure PSPC lipid bilayer at different temperatures, density profiles of the headgroup phosphorous atoms of the pure PSPC bilayer at different temperatures, average order parameter of pure PSPC as a function of temperature, histogram plots for PSPC (310 K), PSPC (330 K), PEG (310 K) and PEG (330 K), instantaneous snapshots of the PTX conformations at 310 K and 330 K for PEGylated multicomponent bilayer membranes.

Table 1: System details for the equilibrium MD simulations. Here, we provide information about the systems, total atoms in the systems, box sizes, temperatures, total simulations time. a*: starting system details for the self-assembly simulation at 360 K. b*: system details after re-solvating the self-assembled system at 360 K.

| Systems | Temperatures (K) | No of PSPC molecules | No of DSPE-PEG molecules | No of counterions | No of water molecules | Total atoms in the systems | Simulation box sizes (nm^3) | Total simulations time (ns) |
|---|---|---|---|---|---|---|---|---|
| Pure PSPC | 360 | 128 | 0 | 0 | 11196 | 50996 | 7x7x13 | 220 |
| | 340 | 128 | 0 | 0 | 11196 | 50996 | 7x7x13 | 220 |
| | 330 | 128 | 0 | 0 | 11196 | 50996 | 7x7x13 | 220 |
| | 320 | 128 | 0 | 0 | 11196 | 50996 | 7x7x13 | 220 |
| | 310 | 128 | 0 | 0 | 11196 | 50996 | 7x7x13 | 220 |
| | 300 | 128 | 0 | 0 | 11196 | 50996 | 7x7x13 | 220 |
| | 280 | 128 | 0 | 0 | 11196 | 50996 | 7x7x13 | 220 |
| Multicomponent bilayer | a*360 | 128 | 8 | 8 | 9230 | 48826 | 15x15x15 | 220 |
| | b*360 | 128 | 8 | 8 | 17248 | 72880 | 10x5x15 | 220 |
| | 340 | 128 | 8 | 8 | 17248 | 72880 | 10x5x15 | 220 |
| | 330 | 128 | 8 | 8 | 17248 | 72880 | 10x5x15 | 220 |
| | 320 | 128 | 8 | 8 | 17248 | 72880 | 10x5x15 | 220 |
| | 310 | 128 | 8 | 8 | 17248 | 72880 | 10x5x15 | 220 |
| | 300 | 128 | 8 | 8 | 17248 | 72880 | 10x5x15 | 220 |
| | 280 | 128 | 8 | 8 | 17248 | 72880 | 10x5x15 | 220 |

Table 2: Provided here details about the systems, temperatures, number of umbrella sampling windows, reaction coordinates, atoms in the systems, simulations time, force constant etc.

| Systems | Temperature (K) | Number of US windows | Maximum, minimum reaction coordinates (nm) | Total atoms in the systems | Each windows simulation time (ns) | Total simulations time (ns) | Force constant (kJ.mol$^{-1}$.nm$^{-2}$) |
|---|---|---|---|---|---|---|---|
| PSPC | 310 | 50 | 5, 0 | 51010 | 10 | 500 | 1000 |
| | 310 | 50 | 5, 0 | 51010 | 10 | 500 | 1000 |
| | 310 | 50 | 5, 0 | 51010 | 10 | 500 | 1000 |
| PSPC | 330 | 50 | 5, 0 | 50989 | 10 | 500 | 1000 |
| | 330 | 50 | 5, 0 | 50989 | 10 | 500 | 1000 |
| | 330 | 50 | 5, 0 | 50989 | 10 | 500 | 1000 |

| | 310 | 50 | 5, 0 | 72861 | 10 | 500 | 1000 |
|---|---|---|---|---|---|---|---|
| PEG | 310 | 50 | 5, 0 | 72861 | 10 | 500 | 1000 |
| | 310 | 50 | 5, 0 | 72861 | 10 | 500 | 1000 |
| | 330 | 50 | 5, 0 | 72888 | 10 | 500 | 1000 |
| PEG | 330 | 50 | 5, 0 | 72888 | 10 | 500 | 1000 |
| | 330 | 50 | 5, 0 | 72888 | 10 | 500 | 1000 |

Table 3: Dissociation barriers and interface barriers have been provided here for pure PSPC and PEGylated bilayers.

| Systems | Temperature (K) | Dissociation barriers (kcal/mol) with errors | Interface barriers (kcal/mol) with errors |
|---|---|---|---|
| PSPC | 310 | 0.04 (0.028) | |
| PEG | 310 | 0.10 (0.036) | |
| PSPC | 330 | 0.30 (0.323) | 0.52 (0.145) |
| PEG | 330 | 0.59 (0.141) | 0.73 (0.032) |

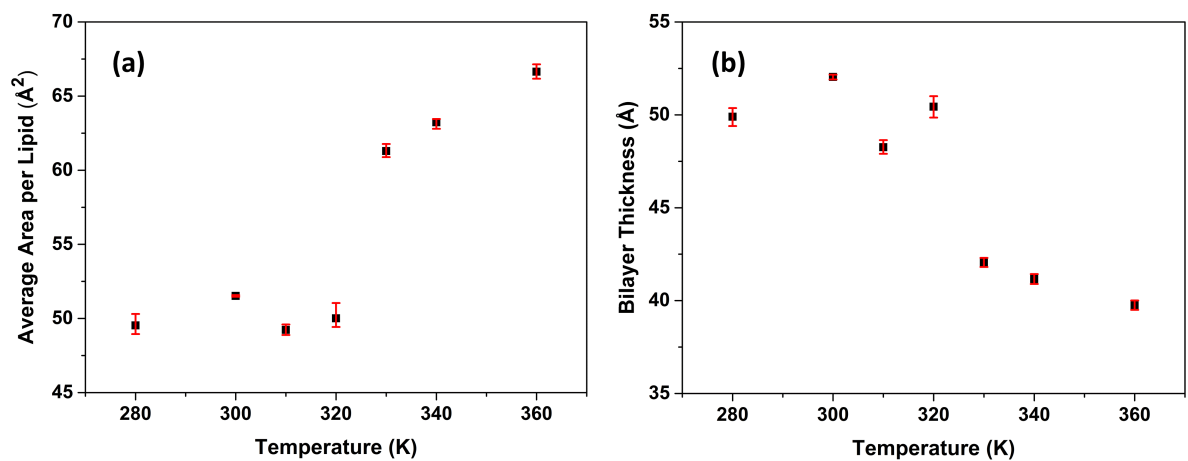

**Figure 1:** (a) Average area per lipid and (b) Bilayer thickness of PSPC as a function of temperature

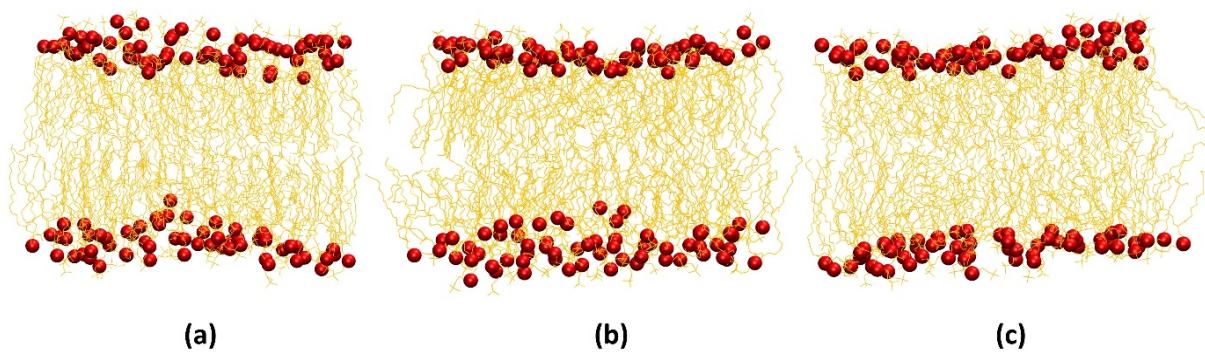

**Figure 2:** Snapshots of PSPC in the liquid crystalline phase at (a) 330 K, (b) 340 K, and (c) 360 K

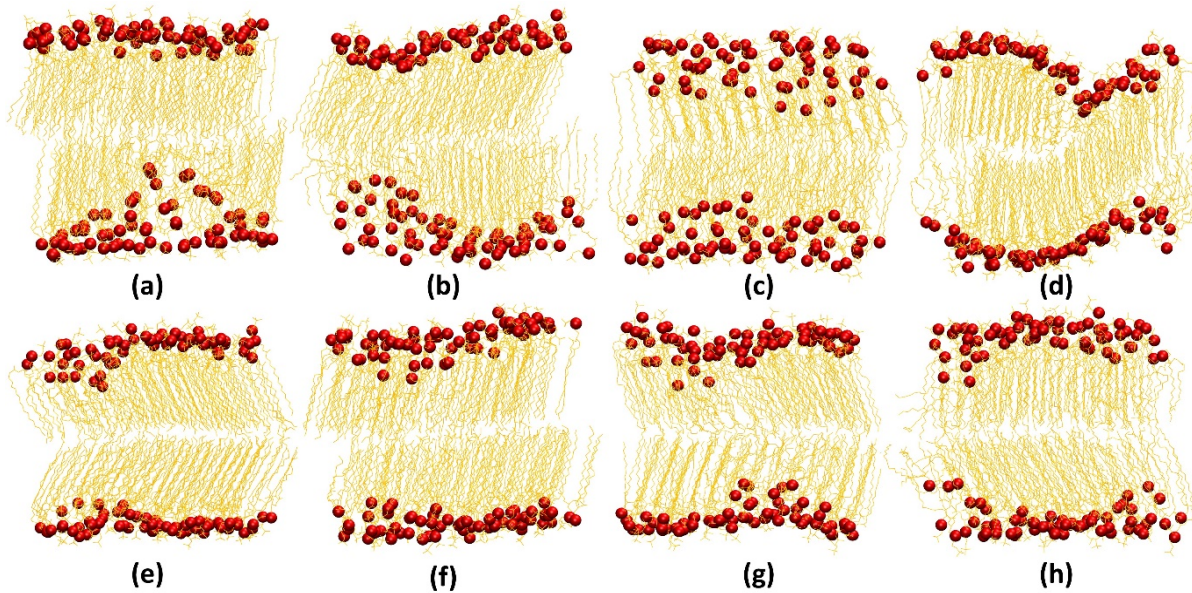

**Figure 3:** Snapshots of PSPC bilayer below phase transition. The structures shown are at different temperatures and viewed from two planes: (a) 280 K x-z plane (b) 280 K y-z plane (c) 300 K x-z plane (d) 300 K y-z plane (e) 310 K x-z plane (f) 310 K y-z plane (g) 320 K x-z plane (h) 320 K y-z plane

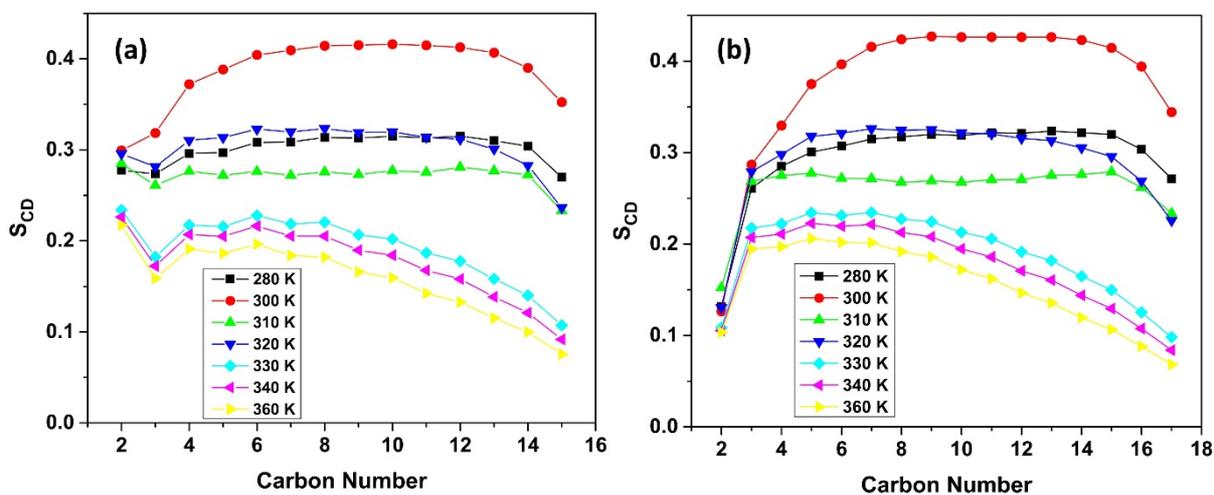

**Figure 4:** Deuterium order parameter of PSPC (a) acyl chain-1 and (b) acyl chain-2 at different temperatures

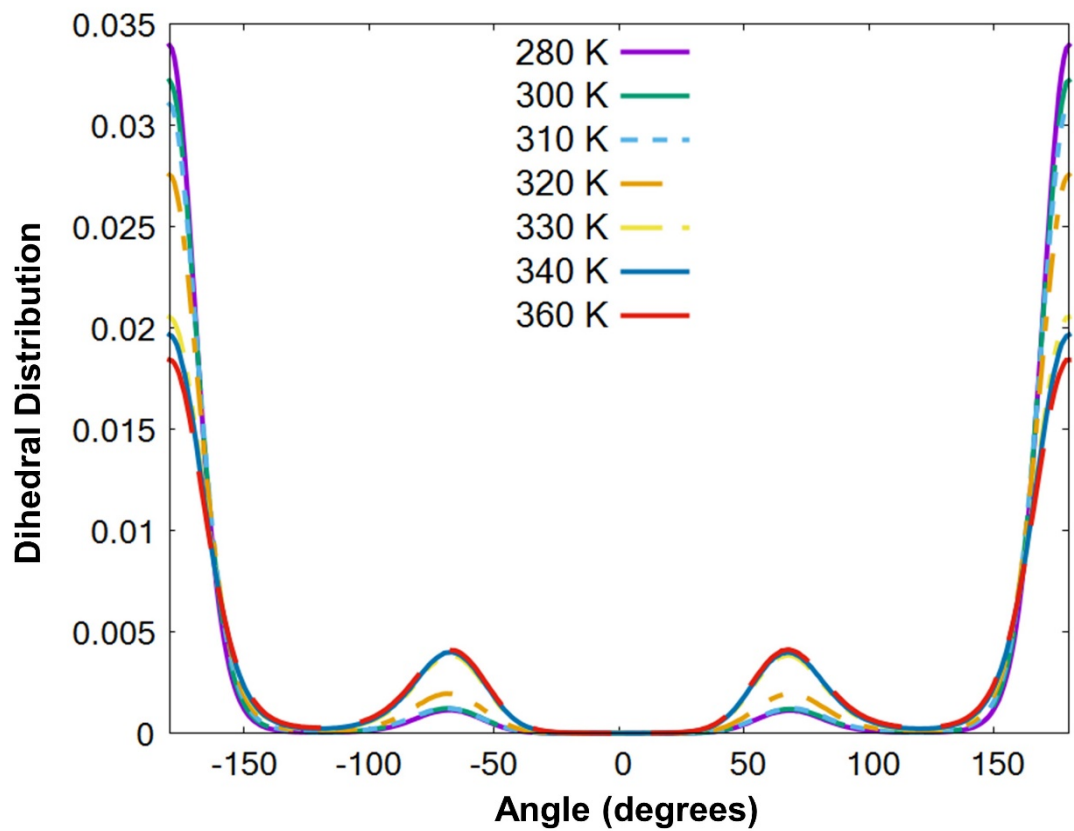

**Figure 5:** Lipid tail dihedral distribution of PSPC at different temperatures.

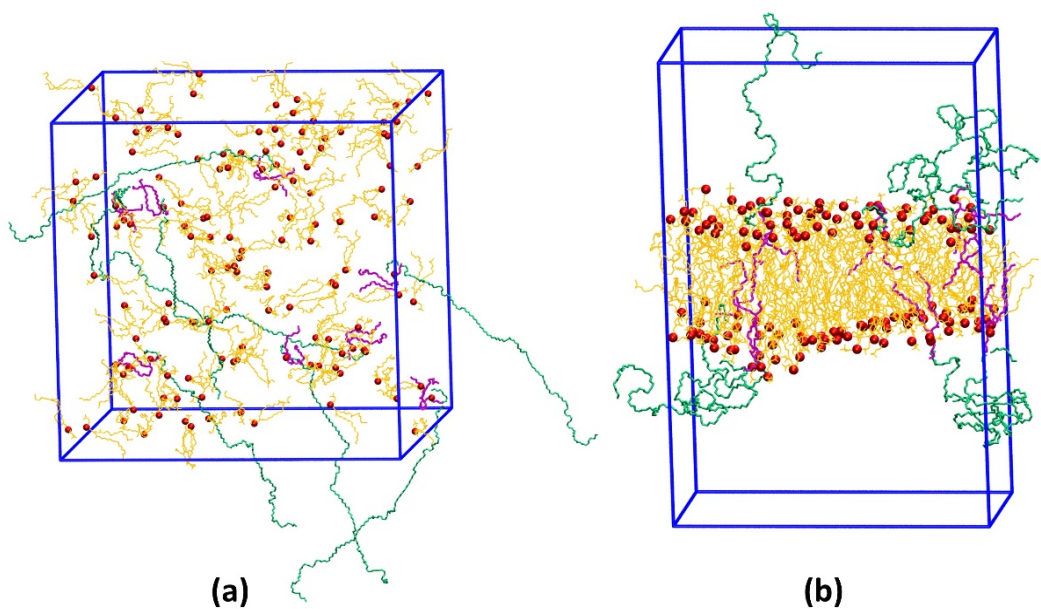

**Figure 6:** (a) Initial and (b) final snapshots of the self-assembly simulation of the multicomponent PSPC and DSPE-PEG system

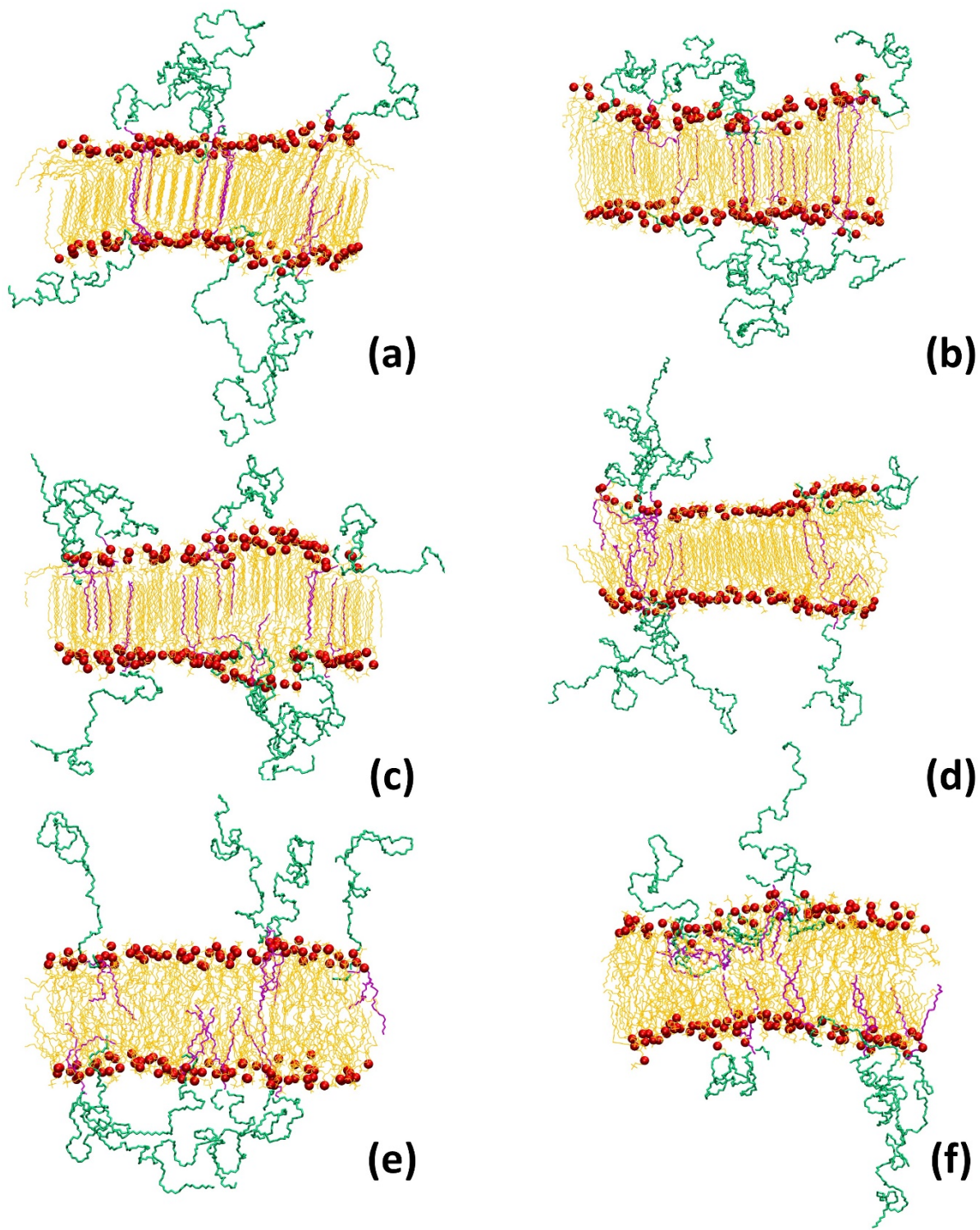

**Figure 7:** Simulation snapshots of the multicomponent PSPC and DSPE-PEG system at (a) 280 K, (b) 300 K, (c) 310 K, (d) 320 K, (e) 330 K, and (f) 340 K.

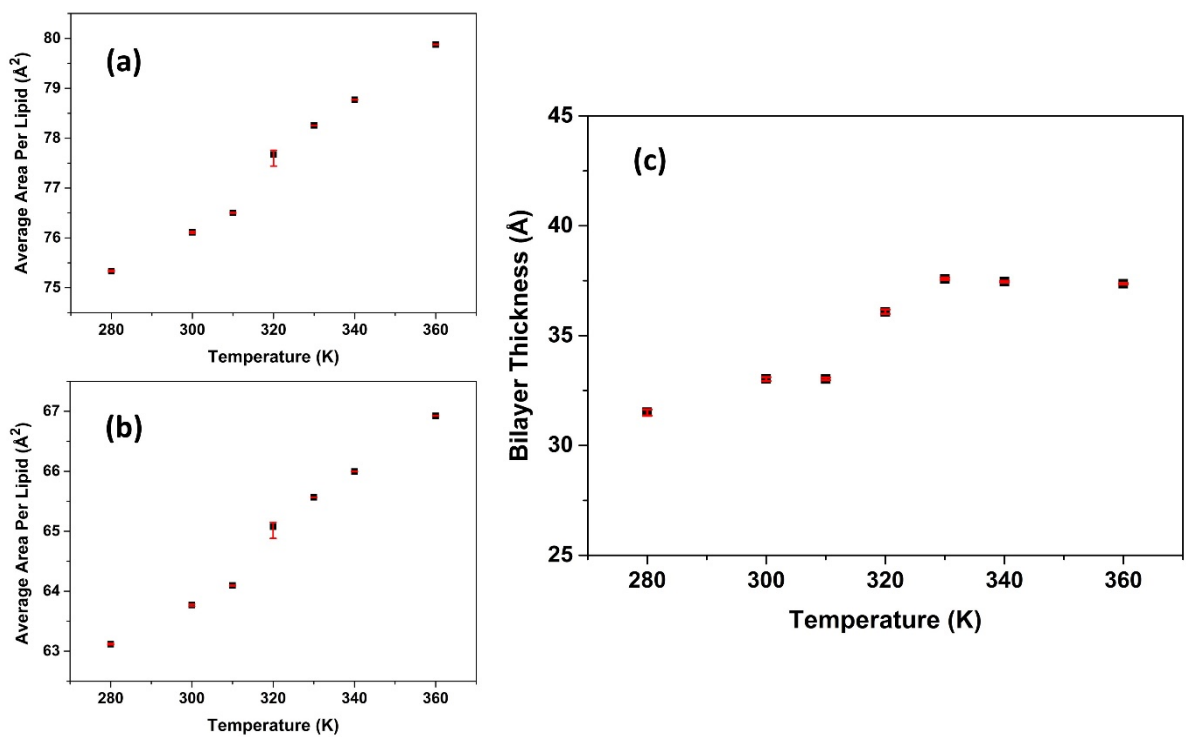

**Figure 8:** Average area per lipid of (a) top leaflet and (b) bottom leaflet, and (c) Bilayer thickness of the multicomponent PSPC and DSPE-PEG bilayer.

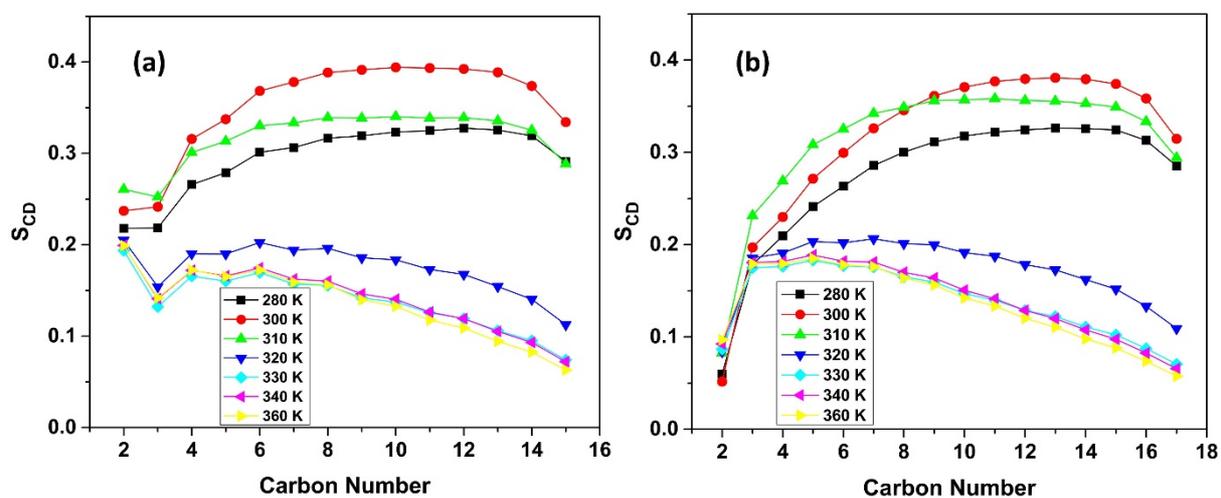

**Figure 9:** Deuterium order parameter of PSPC (a) acyl chain-1 and (b) acyl chain-2 in the multicomponent PSPC and DSPE-PEG system at different temperatures.

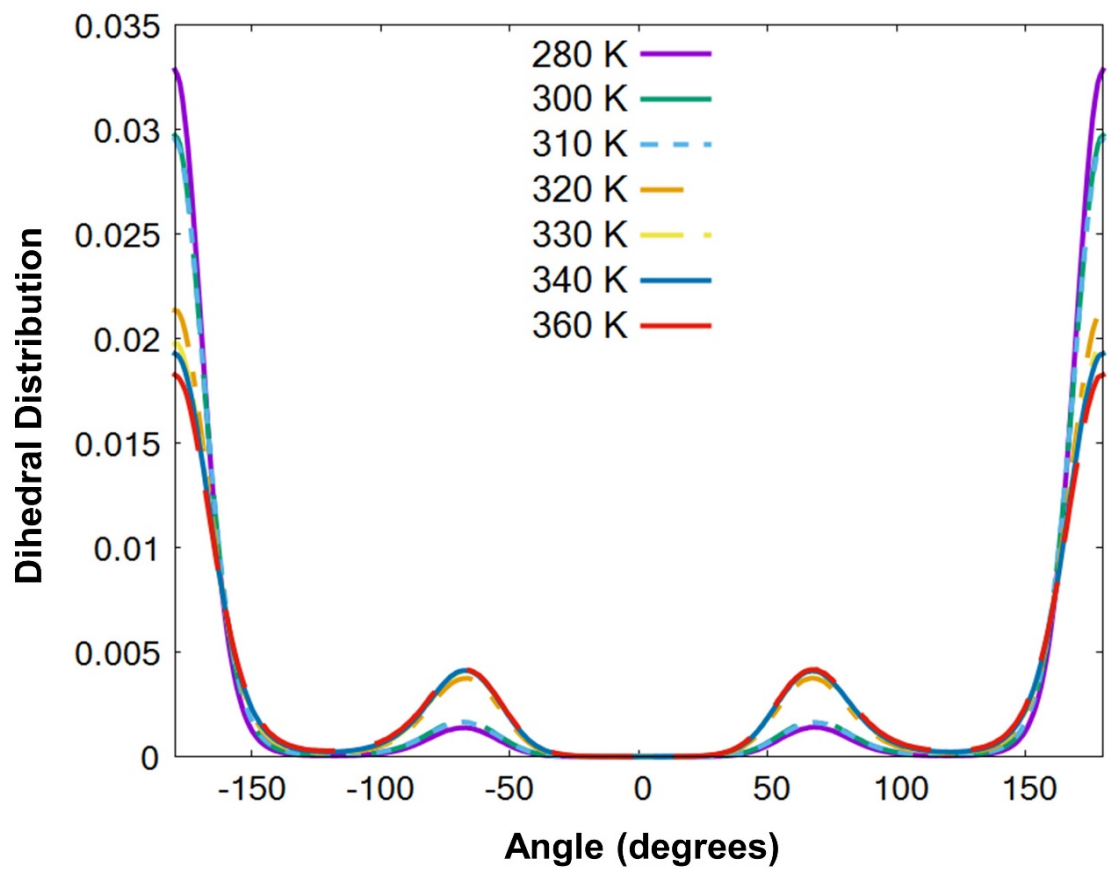

**Figure 10:** Lipid tail dihedral distribution of PSPC in the multicomponent PSPC and DSPE-PEG bilayer at different temperatures.

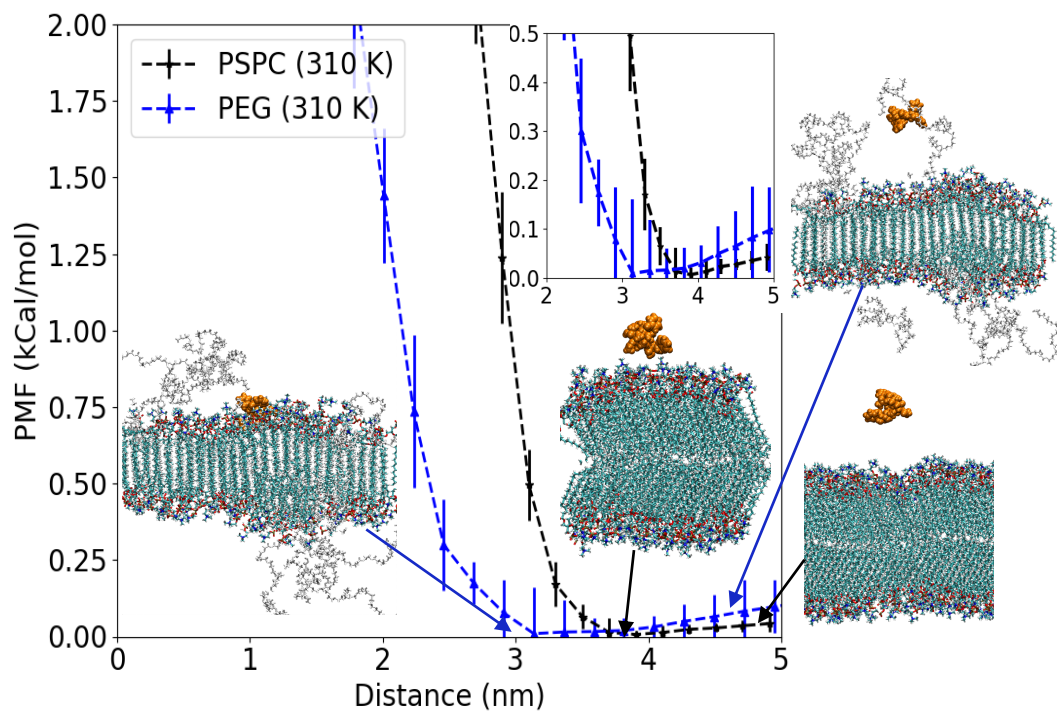

(a)

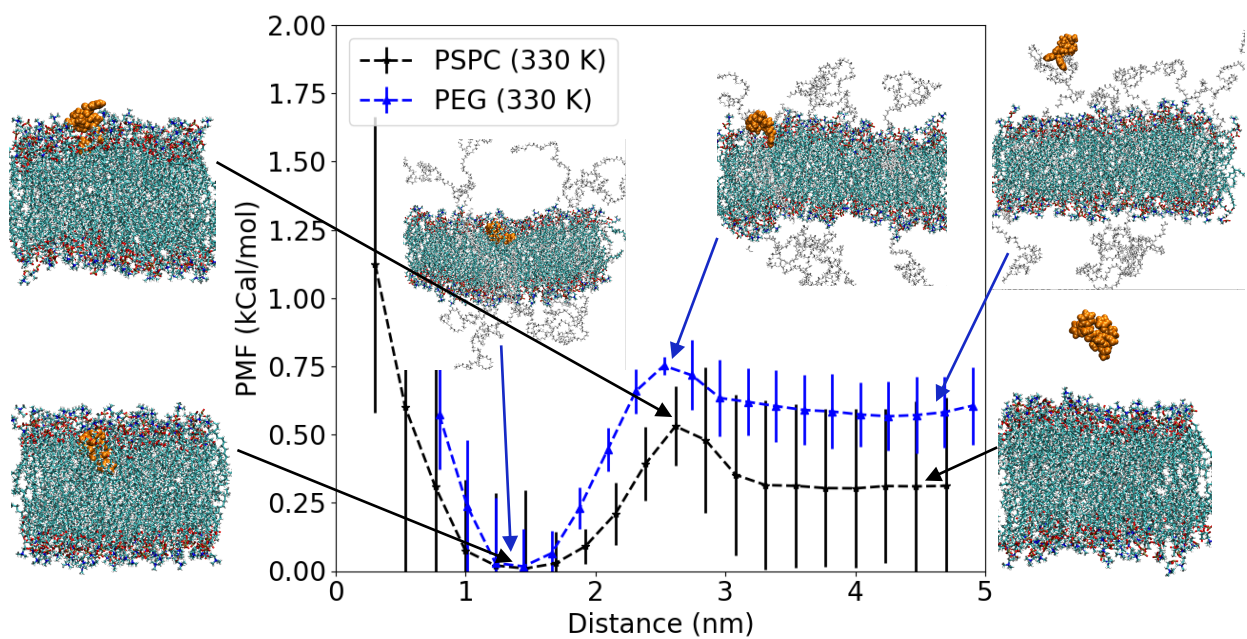

(b)

**Figure 11:** Averaged PMF profiles for (a) pure PSPC (black colour) and PEGylated lipid (blue colour) at 310 K, and (b) pure PSPC (black colour) and PEGylated lipid (blue colour) at 330 K. The inset in Figure (a) shows the PMF profiles near the minimum of the potential well. Instantaneous snapshots show all-atom microscopic pictures at different positions of the potential well for both the systems at two different temperatures. Black and blue arrows point the positions of the instantaneous snapshots with that of the profiles and images. The PTX has been shown in orange colour, DSPE-PEG molecules in grey, lipid bilayer membranes with head and tail groups in green, red colours.